\documentclass[a4paper,10pt]{article}
\pdfoutput=1

\usepackage{jcappub}

\usepackage[style=base, singlelinecheck=off, font=small, labelfont=bf, 
justification=raggedright]{caption}
\usepackage{subfigure}

\usepackage[utf8]{inputenc}

\usepackage{float}
\usepackage{bm}
\usepackage{latexsym}
\usepackage{graphicx}
\usepackage[export]{adjustbox}
\usepackage{color}
\usepackage[dvipsnames]{xcolor}
\usepackage{verbatim}
\usepackage{array}
\usepackage{todonotes}
\numberwithin{equation}{section}
\usepackage{chngcntr} 

\title{Statistical Isotropy Violations in CMB Temperature Anisotropy: Analysis Using Minimal Bipolar Spherical Harmonics}
\author[1,2]{Dipanshu,} 
\emailAdd{garg.dipanshu@students.iiserpune.ac.in} 
\affiliation [1]{Department of Physics, Indian Institute of Science Education and Research, Pune 411008, India}
\affiliation [2]{Raman Research Institute, Bangalore 560080, India}

\author[3]{Akashdeep Karan } 
\emailAdd{akashdeepkaran@students.iisertirupati.ac.in} 
\affiliation [3]{Department of Physics, Indian Institute of Science Education and Research, Tirupati 517507, India}

\author[1,2,4]{Tarun Souradeep}
\emailAdd{tarun@rri.res.in}
\affiliation [4]{Inter University Centre for Astronomy and Astrophysics, Post Bag 4, Ganeshkhind, Pune-411007, India}

\date{\today}
\abstract{ In this work, we present a follow-up to our previous work where the concept of minimal Bipolar spherical harmonics (mBipoSH) functions was introduced as a natural basis for studying angular correlations in the non-statistical isotropic (nSI) Cosmic Microwave Background (CMB) sky. In this study, we extend the formalism of mBipoSH functions and apply it to analyze two sources of statistical isotropy (SI) violation- cosmic hemispherical asymmetry (CHA) and non-circular (NC) beam shapes. We present the analytical expression for mBipoSH functions and plot the corresponding angular correlation functions for these cases, addressing the $L=1$ and $L=2$ multipole statistical isotropy violations within the BipoSH basis. Our results confirm the effectiveness of the proposed approach, as it successfully captures and explains the observed angular correlations in the CMB sky.}

\begin{document}

\maketitle

\section{Introduction }
    The assumption of homogeneity and isotropy of the universe plays an, important role in modern cosmology studies. These assumptions have been tested using various cosmological probes, including the precise measurements of cosmic microwave background (CMB) temperature anisotropy data. Current observations suggest that the anisotropies in the temperature variation of the CMB are in good agreement with them being Gaussian. In such cases, it is possible to fully describe the information encoded in the CMB temperature field using two-point correlation functions. These functions quantify the statistical relationship between the temperature anisotropies at two different points in the sky and can be used to study the statistical properties of the CMB data. 

    Statistical isotropy (SI) refers to the rotational invariance of n-point correlations in a given field. The assumption of SI in CMB data has been under heavy scrutiny after the data from the Wilkinson Microwave Anisotropy Probe (WMAP) and Planck satellite has revealed significant deviations from SI. The measured violation of SI could be a genuine cosmological effect, a residual foreground, a statistical coincidence, a systematic error in the experiment, or a spurious result of inadequate data processing. 

    To extract more information from the rich present and future CMB maps, it is important to design relevant mathematical formalisms to study the anomalies in statistical isotropy. Bipolar Spherical Harmonics (BipoSH) provides an elegant and general formalism to represent the two-point correlation function for a Gaussian distributed field on a 2-dimensional sphere \cite{Hajian_2003}. BipoSH forms a complete and orthogonal basis to study the non-statistical isotropy (nSI) of the covariance matrix. The BipoSH basis has been explored across various known cases of SI violations in CMB data, such as Doppler Boost, Cosmic Hemispherical Asymmetry, Weak Lensing, and several instrumental effects like NC beam. \cite{Hajian_2003, PhysRevD.70.103002, PhysRevD.81.083012, PhysRevD.83.027301, PhysRevD.91.043501, PhysRevD.85.023010, PhysRevD.89.083005, Saha:2021bay, Mukherjee:2013kga, Mukherjee_2016, hajian2004cosmic, Hajian_2004, Mitra_2004, Souradeep_2004,PhysRevD.81.083008}. It is also used in the analysis of the CMB data by the Planck satellite \cite{ade2016planck}, and it is found to be a useful tool for studying the violation of SI in the CMB data. 

    In our previous article \cite{Dipanshu_2023}, we presented an irreducible representation of the BipoSH basis. Using a mathematical framework develoved by Manakov et al. \cite{Manakov_1996}, we demonstrated that this transformation of the BipoSH basis produces a generalized set of functions that can be used to represent correlations in angular separation space. 

    The article is organized as follows. In Section \ref{BipoSH}, we provide an overview of the mBipoSH formalism. This section explores the reduction of the BipoSH basis and its representation, in terms of a new set of real space mBipoSH angular correlation functions.  Section \ref{cases} outlines the process for constructing mBipoSH angular correlation functions for the analysis of CMB anomalies, with a particular focus on the $(L=1)$ Dipole and $(L=2)$ Quadrupole anisotropies, which correspond to cosmic hemispherical asymmetry and non-circular beam nSI effects. Section \ref{Conclusion} marks the conclusion of our findings from the mBipoSH angular correlation functions.

\section{mBipoSH Angular Correlation Functions}
\label{BipoSH}

BipoSH represents a mathematical framework designed to quantify deviations from SI in the temperature or polarisation field data of the CMB sky maps. BipoSH is essentially formed by taking the tensor product of two spherical harmonics basis. This mathematical operation results in the most general basis in the $\mathbf{S}^2 \times \mathbf{S}^2$ space, providing the necessary bidirectional framework for analyzing the CMB two-point function ins spherical harmonic space.

In the context of CMB maps, the BipoSH basis can be expressed as $[{Y_{l_{1}}\left(\hat{n}_{1}\right) \otimes Y_{l_{2}}\left(\hat{n}_{2}\right)}]_{L M}$, where $L$ signifies the harmonic decomposition concerning SI violation in correlation function, and $l$ signifies the harmonic decomposition for the angular scale decomposition at the map level.The most general two-point correlation function for a random field defined on the two-sphere can be obtained in terms of BipoSH basis as

 \begin{equation}\label{eq1}
      \langle \Delta T({\hat n_1}) \Delta T({\hat n_2}) \rangle =  C\left(\hat{n}_{1}, \hat{n}_{2}\right)=\sum_{L, M, l_{1}, l_{2} } A_{l_{1} l_{2}}^{L M}
        \left\{Y_{l_{1}}\left(\hat{n}_{1}\right) \otimes Y_{l_{2}}\left(\hat{n}_{2}\right)\right\}_{L M}.
        \end{equation}
        
         The BipoSH coefficients $A_{l_{1} l_{2}}^{L M}$ and BipoSH functions $[{Y_{l_{1}}\left(\hat{n}_{1}\right) \otimes Y_{l_{2}}\left(\hat{n}_{2}\right)}]_{L M}$ are the building blocks of the two-point correlation function, describing the dependence of the CMB temperature field on the two directions $\hat{n}_1$ and $\hat{n}_2$. The Clebsch-Gordan coefficients $C_{l_{1} m_{1} l_{2}-m_{2}}^{L M}$ defines the product space the two spherical harmonics to form the BipoSH functions as,
 \begin{equation}
      \{Y_{l_{1}}\left(\hat{n}_{1}) \otimes Y_{l_{2}}\left(\hat{n}_{2}\right)\right\}_{L M}=\sum_{m_{1} m_{2}} C_{l_{1} m_{1} l_{2}-m_{2}}^{L M} Y_{l_1,m_1}(\hat n_1) Y_{l_2,m_2}(\hat n_2),
        \end{equation}
The various BipoSH coefficients capture SI violation features in CMB data. The BipoSH coefficient $A^{00}_{ll}$, corresponding to $L = 0$ in the BipoSH basis expansion represents the isotropic component in the CMB sky, which is related to the well-studied power spectrum of the CMB, denoted as $C_l$. On the other hand, when $L \neq 0$, the BipoSH coefficients represent the corresponding bipolar multipoles of nSI effects in the CMB sky. This establishes BipoSH coefficients as a natural extension of the CMB angular power spectrum, widely studied to characterize various sources of non-statistical isotropy effects and systematically investigate nSI in CMB full sky maps.


The BipoSH formalism reveals signatures of SI violations in terms of the harmonic scale of the map, denoted by $l$. In this section, we introduce a novel extension of the BipoSH basis to capture the SI violation features in angular separation $\theta$ dependent real space correlation functions\cite{Dipanshu_2023, Manakov_1996, VMK}.
To establish the mathematical formalism for performing the real space transformation of the Bipolar functions, capturing the nSI features, we utilize the special reduction formulae studied by Manakov et al \cite{Manakov_1996}. BipoSH as defined in the above section, represent tensors of rank $L$. Following general mathematical principles, a tensor with a rank of $L$ should be constructed using $L$ vectors, each formed from its arguments. Utilizing this approach, we perform the reduction of the BipoSH basis. This reduction process of BipoSH functions yields $L$ functions, which, when expressed in angular separation $\theta$ space, manifest as measurable quantities, dubbed as mBipoSH angular correlation functions, that  generalise the SI angular correlation function $C(\theta)$ \cite{Dipanshu_2023}.

Many of the SI violation studies focus on the nSI effects that are relevant only at very low multipoles of CMB anisotropy. In such instances, the reduced bipolar harmonic basis provides an efficient way of studying the nSI effects. It can be seen that in the bipolar harmonic basis, the nSI violation feature could be spread over the entire range of \( l \) from zero to infinity, whereas reduced bipolar harmonics limit our study in real space theta function, having a range of 0 to 180 degrees. Hence, it is crucial to study the nSI features using the reduced bipolar harmonics basis.

Utilizing the reduction mechanism of the bipolar harmonics basis discussed in the \nameref{appendix}, the most general two-point correlation function from Eq. \eqref{eq1} and Eq. \eqref{eq4} can be expressed in the form of the reduced bipolar harmonic basis as

  \begin{equation}\label{eq8}
        C\left(\hat{n}_{1}, \hat{n}_{2}\right)=\sum_{ L, M, l_{1}, l_{2}} A_{l_{1} l_{2}}^{L M}
        \sum_{\lambda=\lambda_p}^{L} a_{\lambda}(l_1,l_2,L,\cos\theta)  Y^{\lambda,L+\lambda_p-\lambda}_{LM} (\hat n_1, \hat n_2).
        \end{equation}

The above equation can be easily recast to another form in case of even parity by exchanging the summations as follows
 \begin{equation}\label{eq9}
      C(\hat n_1,\hat n_2) = \sum_{L, M} \sum_{\lambda=0}^{L} \left[\sum_{l_1,l_2} A^{LM}_{l_1l_2}  a_{\lambda}(l_1,l_2,L,\cos\theta) \right] Y^{\lambda,L-\lambda}_{LM} (\hat n_1, \hat n_2).
  \end{equation}

Now, it is interesting to note that the equation inside the square exhibits different \(L+1\) values for a given multipole \(L\) in our newly defined reduced mBipoSH basis. These values correspond to the theta-dependent correlation functions expanded in the general nSI sky, referred to as the Minimal Bipolar (mBipoSH) correlation functions. This set of angular correlation functions can most generally encapsulate nSI features in the sky maps.

The proposed method offers a comprehensive approach to non-statistical isotropic (nSI) signatures in the CMB sky, akin to the BipoSH basis. In an observed CMB sky map, the temperature field map may exhibit multiple sources of nSI signatures. The mBipoSH angular correlation functions establish a set of isotropic, angular separation ($\theta$) dependent functions in a general SI violated CMB sky map. It is noteworthy that, for \(L = 0\), the only  mBipoSH angular correlation function corresponds to the isotropic two-point correlation function, widely known as the CMB angular correlation function. The angular correlation function, $C(\theta)$ has been extensively studied in cosmological literature, serving as a rich source of information.

To summarize, we deduce that the mBipoSH angular correlation functions for any given multipole \(L\) in a nSI CMB sky can be expressed as a product of BipoSH coefficients and newly defined \(a_{\lambda}\) coefficients as

\begin{equation}\label{qq}
\alpha^{L,M}_{\lambda}(\cos \theta) = \sum_{l_{1} l_{2}} A_{l_1 l_2}^{L M} a_{\lambda}(l_1,l_2,L,\cos\theta),
\end{equation}

Utilizing the definition of bipolar spherical harmonics in terms of spherical harmonics, this expression can be further rephrased as

\begin{equation}
\alpha^{L,M}_{\lambda}(\cos \theta) = \sum_{l_{1} l_{2}} \sum_{m_{1} m_{2}} a_{l_{1} m_{1}} a_{l_{2} m_{2}}^{*}(-1)^{m_{2}} C_{l_{1} m_{1} l_{2}-m_{2}}^{L M} a_{\lambda}(l_1,l_2,L,\cos\theta).
\end{equation}

Therefore, mBipoSH angular basis offers a real-space counterpart to the spherical harmonics basis investigated in CMB studies. Given that \(a_{lm}\) coefficients are derived from CMB temperature maps containing real-space temperature fluctuations information, it becomes imperative to examine the CMB sky in real space. In our previous paper, we focused on the predominant nSI effect of the Doppler boost in simulated CMB sky maps. In this study, we extend our exploration to encompass additional cases of CMB SI violation.  

\section{mBipoSH correlation functions for non-SI CMB features}
\label{cases}             

\subsection{Cosmic Hemispherical Asymmetry (CHA)}
Much like the Doppler boost nSI effect, the Cosmic Hemispherical Asymmetry (CHA) stands out as a well-discussed nSI effect in cosmic microwave background sky maps, manifesting as an \(L=1\) dipole anisotropy. CHA is studied as a dipole modulation of a statistically isotropic (SI) CMB sky. Similar to the analysis of the Doppler boost, the study of CHA involves exploring the \(l, (l+1)\) correlations in the BipoSH basis. The CMB temperature anisotropy field, modulated by the dipole in the direction \(\hat{n}\), is mathematically expressed as

$$
\Delta T(\hat{n})=(1+A \hat{p} \cdot \hat{n}) \Delta T^{S I}(\hat{n}) .
$$

Here, $\Delta T^{S I}(\hat{n})$ is the SI CMB temperature field and $\Delta T(\hat{n})$ is the observed nSI temperature field. The direction of the modulation dipole is specified by $\hat{p} \equiv\left(\theta_p, \phi_p\right)$ with $A$ representing the amplitude of the dipole modulation.

In the context of cosmic hemispherical asymmetry, the amplitude of dipole modulation is studied to exhibit scale-dependent characteristics. In this scenario, the scale-dependent modulation field is described using the amplitude function \(A(l)\) and the dipole direction vector \( \hat{p} \), expressed in terms of the spherical harmonic basis as
$$
A(l) \hat{p} \cdot \hat{n}=\sum_{N=-1}^1 m_{1 N}(l) Y_{L=1, N}(\hat{n}).
$$
The equations that describe the amplitude \(A\) and direction \((\theta_p, \phi_p)\) in terms of the spherical harmonic coefficients \(m_{10}\), \(m_{11}^r\), and \(m_{11}^i\) are given as

$$
\begin{aligned}
& A(l)=\sqrt{\frac{3}{4 \pi}} \sqrt{m_{10}(l)^2+2 m_{11}^r(l)^2+2 m_{11}^i(l)^2}, \quad \theta_p=\cos ^{-1}\left[\frac{m_{10}(l)}{A(l)} \sqrt{\frac{3}{4 \pi}}\right] \\
& \text { and } \phi_p=-\tan ^{-1}\left[\frac{m_{11}^i(l)}{m_{11}^r(l)}\right] .
\end{aligned}
$$

\noindent In this study, the dipole modulation amplitude \(A(l)\) is modeled using the power-law model as investigated in Shaikh et al \cite{Shaikh_2019}. The power-law model expresses the dipole modulation amplitude \(A\) as
$$
A(l)=A\left(l_p\right)\left(\frac{l}{l_p}\right)^\alpha,
$$
Where \(A\left(l_p\right)\) represents the amplitude at a selected multipole \(l_p\), the power-law dependence of the amplitude can be expressed in the form of spherical harmonic coefficients of the dipole modulation field as
$$
m_{1 N}(l)=m_{1 N}\left(l_p\right)\left(\frac{l}{l_p}\right)^\alpha
$$
The CHA effect can be effectively modeled by capturing the correlations between $l$ and $l+1$ in the BipoSH  basis. Using the mathematical framework  defined above for CHA, the BipoSH coefficients can be written as

$$
A_{l l+1}^{1 N}\left(l_p, \alpha\right)=m_{1 N}\left(l_p\right) G_{l l+1}^1,
$$
where $G_{l l+1}^1$ is  the shape factor for the cosmic hemispherical asymmetry nSI signal. Specifically, here
$$
G_{l l+1}^1 \equiv \frac{\Pi_{l l+1}}{\sqrt{12 \pi}}\left[\left(\frac{l}{l_p}\right)^\alpha C_l+\left(\frac{l+1}{l_p}\right)^\alpha C_{l+1}\right] C_{l 0 l+10}^{10} .
$$
The most general correlation function for a nSI map can be expressed in terms of mBipoSH angular functions as

\begin{equation}\label{eq9}
      C(\hat n_1,\hat n_2) = \sum_{L, M} \sum_{\lambda=0}^{L} \alpha^{L,M}_{\lambda}(\cos \theta) Y^{\lambda,L-\lambda}_{LM} (\hat n_1, \hat n_2).
  \end{equation}

The nSI CMB sky map corresponding to cosmic hemispherical asymmetry effect will include the dipole signal $(L=1)$ in addition to the standard isotropic (SI) contribution at $L=0$. This can be expanded as follows
  
\begin{equation}\label{eq9}
      C(\hat n_1,\hat n_2) = C_{SI}(\cos \theta) + \sum_{M=(-1,0,1)} \sum_{\lambda=0}^{1} \alpha^{1,M}_{\lambda}(\cos \theta) Y^{\lambda,1-\lambda}_{1M} (\hat n_1, \hat n_2).
  \end{equation}
  As the direction of dipole modulation is modeled as $'p,'$ designating this as our primary direction in the BipoSH basis, it follows that only the $M = 0$ component will persist, having  other $M$ components null. Employing these, we obtain correlation function having SI and nSI part as
\begin{equation}\label{eq9}
      C(\hat n_1,\hat n_2) = C_{SI}(\cos \theta) + \alpha^{1,0}_{0}(\cos \theta) [n_1]_{1, 0}
 +  \alpha^{1,0}_{1}(\cos \theta)  [n_2]_{1, 0}.
  \end{equation}

  \begin{figure}[htbp]
  \centering
  \includegraphics[width=0.5\textwidth]{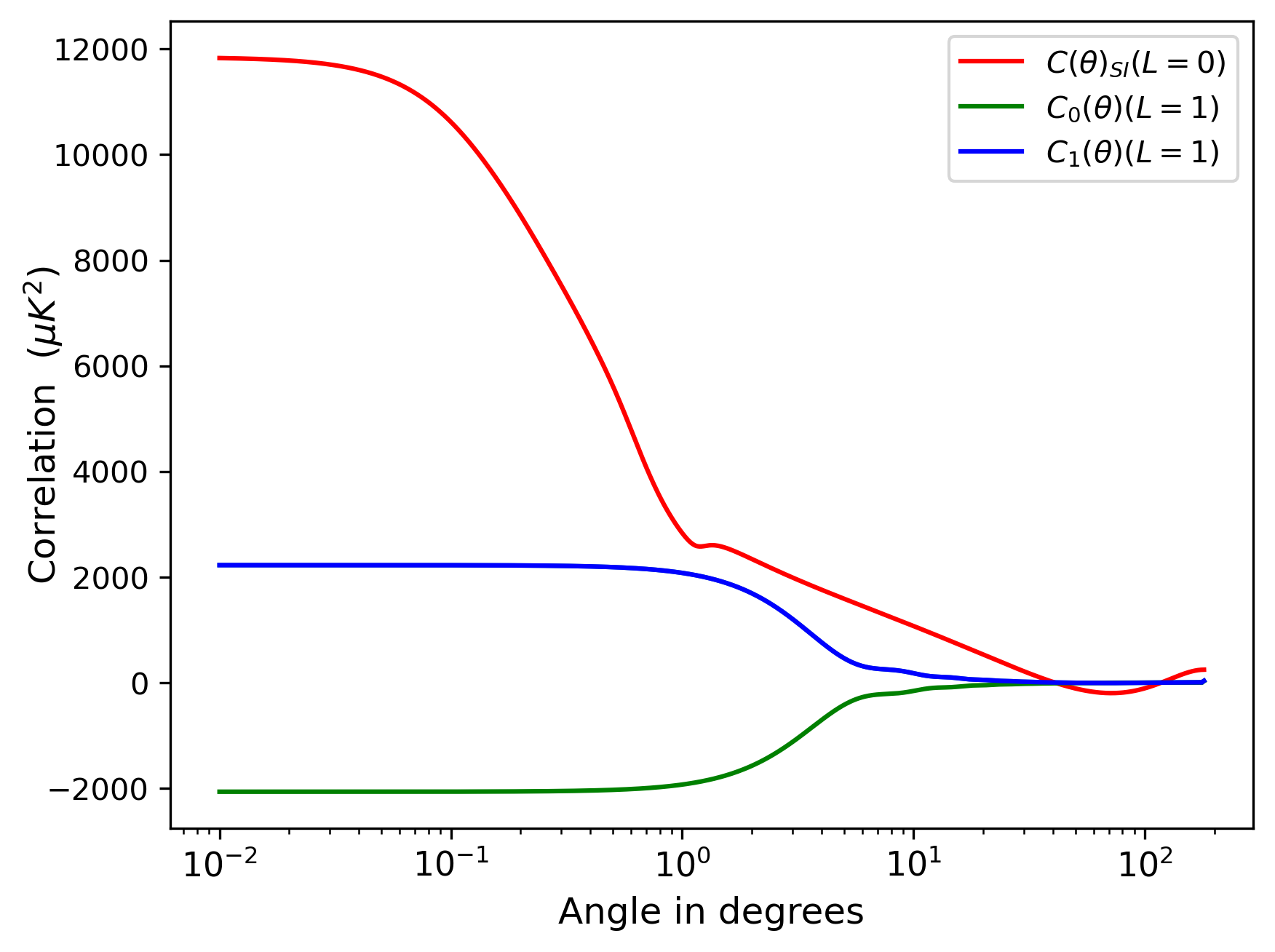}\hfill
  \includegraphics[width=0.5\textwidth]{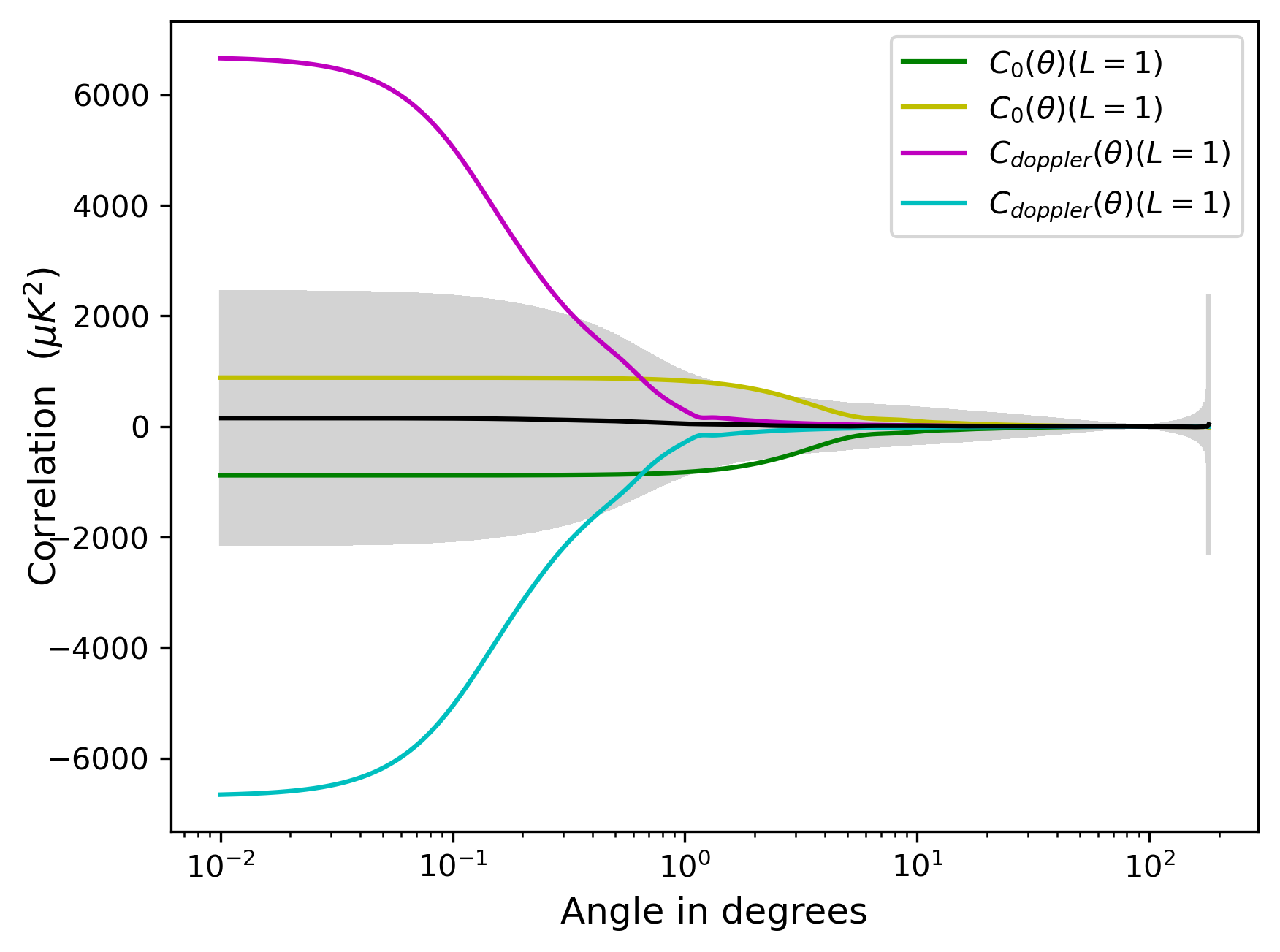}
  \caption{The left panel displays the mBipoSH correlation functions for the (nSI effect) CHA, namely $C_0(\theta)$ and $C_1(\theta)$ with parameters along with the SI correlation function. These plots are generated assuming the angular power spectrum $C_l^{TT}$ computed for the best-fit $\Lambda$CDM model parameters using CAMB \citep{2011ascl.soft02026L}. The right panel presents the cosmic variance error bar for mBipoSH correlation functions corresponding to the cosmic hemispherical asymmetry effect, using 1000 simulated cosmic hemispherical asymmetry maps. We used the CoNIGS code to generate the simulated nSI maps for the cosmic hemispherical asymmetry (\cite{Mukherjee:2013kga})}
\label{fig:e1}
  \label{fig:both_images}
\end{figure}   
  Calculating the mBipoSH functions using the equation \eqref{eq9} and the relation $[n]_{1, 0} = \hat{n} \cdot \hat{p}$, the correlation function can be expanded as 
\begin{equation}\label{eq18}
\begin{split}
            C_{nSI}\left(\hat{n}_{1}, \hat{n}_{2}\right)  = C_0 (\cos\theta)  \hat{n}_1. \hat{p} + C_1 (\cos\theta)  \hat{n}_2. \hat{p}
            \end{split}
        \end{equation}

        Here, $C_0 (\cos\theta)$ and $C_1 (\cos\theta)$ represent the set of two isotropic angular correlation functions corresponding to dipole anisotropy associated with CHA. These functions can be expressed as

 \begin{equation}
    C_0(\theta)  = \sum_{l}m_{1 N}\left(l_p\right) G_{l l+1}^1  \left[\frac{-\sqrt{3}}{4 \pi \sqrt{l+1}}(-1)^{l}P_{l}^{(1)}(\cos \theta)\right]
                \end{equation}
 \begin{equation}
  C_1(\theta)   = \sum_{l} m_{1 N}\left(l_p\right) G_{l l+1}^1\left[\frac{-\sqrt{3}}{4 \pi \sqrt{l+1}}(-1)^{l+1}P_{l+1}^{(1)}(\cos \theta)\right].
                \end{equation}
  In the depicted figure, it becomes evident that the cosmic hemispherical effect exhibits minimal impact at extremely small angles when $L=1$, especially in comparison to the pronounced influence of the Doppler boost effect. However, as the angles increase, the cosmic hemispherical effect progressively becomes more dominant. Analysis of the error bar plot leads us to the conclusion that the signal strength of the cosmic hemispherical anomaly resides within the confines of the cosmic variance bars delineated in the Statistical Isotropic plots. Consequently, drawing a decisive inference regarding the presence of the cosmic hemispherical effect versus random realizations of the CMB sky is inconclusive at this juncture.    

\subsection{Non Circular Beam}
The observed cosmic microwave background anisotropy on the sky results from convolving the intrinsic cosmological CMB signal with the instrumental beam response function. Inevitable deviations from circular symmetry in instrumental beams, coupled with specific scan strategies, can lead to detectable violations of SI in CMB experiments. Exploring this nSI effect in CMB data analysis involves focusing on the \(L=2\) quadrupole effect as measured in WMAP-7 data \cite{weiland2011seven,das2016statistical}. This is particularly relevant when modeling the anisotropic beam as an elliptical case. This approach enables a thorough investigation of how deviations from circular symmetry in instrumental beams contribute to SI violations in the CMB data. 

To quantify SI violation for the NC beam case, studying its effect in the BipoSH basis leads to expansion coefficients known as beam-BipoSH coefficients (\(B^{LM}_{l_1l_2}\)). For an ideal circularly symmetric beam, only the \(L = 0\) beam-BipoSH coefficients are non-vanishing. However, the breakdown of circular symmetry further induces \(L \neq 0\) modes in these coefficients. 

The observed CMB temperature results from the convolution of the actual underlying CMB temperature with the instrument beam can be expressed as
$$
\widetilde{\Delta T}\left(\hat{n}_1\right)=\int d \Omega_{\hat{n}_2} B\left(\hat{n}_1, \hat{n}_2\right) \Delta T\left(\hat{n}_2\right)
$$.

Here, $\Delta T\left(\hat{n}_2\right)$ is the underlying true sky temperature along $\hat{n}_2$ and $\widetilde{\Delta T}\left(\hat{n}_1\right)$ is the temperature measured along the direction $\hat{n}_1$ .  The factor $B\left(\hat{n}_1, \hat{n}_2\right)$ is known as the beam response function and gives the sensitivity of the detector around the pointing direction, $\hat{n}_1$.

The beam response function can be expanded in terms of BipoSH basis as
\begin{equation}
B\left(\hat{n}_1, \hat{n}_2\right)=\sum_{l_1 l_2 L M} B_{l_1 l_2}^{L M}  Y^{l_1 l_2}_{L M},
\end{equation}
where, $ B_{l_1 l_2}^{L M} $ represents the beam-BipoSH coefficients. The BipoSH coefficients that capture the non-circular (NC) nSI effect from the observed two-point temperature correlation function can be expressed in terms of the Beam-BipoSH coefficients as \cite{pant2015estimating}
\begin{equation} \label{beam biposh w/o pt}
\tilde{A}_{l_1 l_2}^{L_1 M_1}=\sum_{l L M L^{\prime} M^{\prime}} C_l B_{l_1 l}^{L M} B_{l_2 l}^{L^{\prime} M^{\prime}}(-1)^{l_1+L^{\prime}-L_1} \sqrt{(2 L+1)\left(2 L^{\prime}+1\right)} C_{L M L^{\prime} M^{\prime}}^{L_1 M_1}\left\{\begin{array}{ccc}
l & l_1 & L \\
L_1 & L^{\prime} & l_2
\end{array}\right\}    
\end{equation}

The beam-BipoSH's behaviour doesn't just depend on NC-beam harmonics. It also relies on how we scan the sky, which is described by $\rho(\hat{n})$ for different pointing directions ($\hat{n} \equiv (\theta, \phi)$). The formula for beam-BipoSH in equation\eqref{beam biposh w/o pt} can be dealt with more easily when the scanning pattern makes $\rho(\hat{n})$ stay the same. We call this kind of scanning pattern a 'parallel-transport' (PT) scan, as mentioned in \cite{souradeep2001window}. This means the beam's orientation stays constant compared to the local longitude wherever you look in the sky. This specific scanning method is useful when dealing with a uniform sky, leading to $M = 0, M' = 0, M_1 = 0$. In that case, the BipoSH coefficients can be written as

\begin{equation}\label{beam BipoSH PT}
\tilde{A}_{l_1 l_2}^{L_1 0}=\sum_{l L L^{\prime}} C_l B_{l_1 l}^{L 0} B_{l_2 l}^{L^{\prime} 0}(-1)^{l_1+L^{\prime}-L_1} \times \sqrt{(2 L+1)\left(2 L^{\prime}+1\right)} C_{L 0 L^{\prime} 0}^{L_1 0}\left\{\begin{array}{ccc}
l & l_1 & L \\
L_1 & L^{\prime} & l_2
\end{array}\right\} .
\end{equation}

The NC beam effect can be computed within the \(L=2\) multipole using the PT scan strategy. This involves correlations of the types \(l\,\ l\) and \(l\, (l+2)\) in harmonic space. By applying this condition to the \eqref{beam BipoSH PT}, the BipoSH coefficients can be expressed as

\begin{equation}
\begin{aligned}
\tilde{A}_{l l}^{20} & =\frac{(-1)^l 2 \sqrt{5} C_l B_{l l}^{00} B_{l l}^{20}}{\left(\prod_l\right)^3 C_{l 0 l 0}^{20}} \\
\tilde{A}_{l l+2}^{20 } & =\frac{\sqrt{5}(-1)^l}{\prod_{l l+2} C_{l0 l+2 0}^{20}} \times\left[\frac{C_{l} B_{l l}^{00} B_{l+2 l}^{20}}{\prod_{l}}+\frac{C_{l+2} B_{l+2 l+2}^{00} B_{l l+2}^{20}}{\prod_{l+2}}\right]
\end{aligned}
\end{equation}
The analytical expression for bema-BipoSH coefficients under parallel transport scan approximation can be given by,
\begin{equation} \label{B_LM beam biposh epression}
B_{l_1 l_2}^{L M}=\delta_{M 0} \frac{2 \pi \prod_{l_1}}{\sqrt{4 \pi}}\left(b_{l_2 2}(\hat{z})+b_{l_2 2}^*(\hat{z})\right)\left[C_{l_1 0 l_2 0}^{L 0} I_{0,2}^{l_1 l_2}+\sum_{m_2 \neq 0}(-1)^{m_2} C_{l_1-m_2 l_2 m_2}^{L 0} I_{m_2, 2}^{l_1 l_2}\right]
\end{equation}
where the $b_{l m}$'s are the spherical harmonic coefficients of the beam response function and,
\begin{equation}
    I_{m_2, m^{\prime}}^{l_1 l_2}=\int d_{m_2 m^{\prime}}^{l_2}(\theta) d_{m_2 0}^{l_1}(\theta) \sin \theta d \theta .
\end{equation}
Here, $d^{l}_{m m'}$'s are Wigner-d functions.

A trivial break of symmetry for a circular symmetric Gaussian function is the Elliptical Gaussian (EG) function. This function is often used as a good approximate for the NC beam. An EG beam with its pointing direction along $\hat{z}$ axis can be expressed as,
\begin{equation}
    B(\hat{z}, \hat{n})=\frac{1}{2 \pi \sigma_1 \sigma_2} \exp \left[-\frac{\theta^2}{2 \sigma^2(\phi)}\right]
\end{equation}
where $\sigma_1$ and $\sigma_2$ are the Gaussian widths along the semi-major and semi-minor axis. The azimuth-dependent beam-width $\sigma^2(\phi)=\left[\sigma_1^2 /\left(1+\epsilon \sin ^2 \phi_1\right)\right]^{1 / 2}$ depends on the non-circularity parameter $\epsilon$. This non-
circularity parameter is can be related to the eccentricity $e$ as $\epsilon=\frac{e^2}{1-e^2}$. 

A more commonly used parameter to define the size of the beam is Full Width Half Maxima, $\theta_{FWHM}$. Which can be related to the semi-major and semi-minor axis using the following equation,
\begin{equation}
    \theta_{FWHM}=\sqrt{8 ln 2 \times \sigma_1 \sigma_2}
\end{equation}

The beam spherical harmonic coefficients of such EG beam is given by\cite{souradeep2001window},
\begin{equation}
    \begin{aligned}
& b_{l m}(\hat{z})=\sqrt{\frac{(2 l+1)(l+m)!}{4 \pi(l-m)!}}\left(l+\frac{1}{2}\right)^{-m} \times \\
& \quad I_{m / 2}\left[\left(l+\frac{1}{2}\right)^2 \frac{\sigma_1^2 e^2}{4}\right] \exp \left[-\left(l+\frac{1}{2}\right)^2 \frac{\sigma_1^2}{2}\left(1-\frac{e^2}{2}\right)\right]
\end{aligned}
\end{equation}
where, $I_{\nu}(x)$ is the modified Bessel function.

Utilizing the formalism established in the preceding section, the correlation function can be expanded in terms of mBipoSH angular correlation functions in a similar manner for this particular case, as
\begin{equation}\label{eq9}
      C(\hat n_1,\hat n_2) = C_{SI}(\cos \theta) + \sum_{\lambda=0}^{2} \alpha^{2,0}_{\lambda}(\cos \theta) Y^{\lambda,2-\lambda}_{20} (\hat n_1, \hat n_2).
  \end{equation}
The equation can be simplied to set of isotropic correlation functions as
  \begin{equation}\label{eq9}
      C(\hat n_1,\hat n_2) = C_{SI}(\cos \theta) + C_{0}(\cos \theta) f_{0}(\hat n_1,\hat n_2)
 +  C_{1}(\cos \theta)  f_{1}(\hat n_1,\hat n_2) + C_{2}(\cos \theta)f_{2}(\hat n_1,\hat n_2).
  \end{equation}
where $C_{0}(\cos \theta)$, $C_{1}(\cos \theta)$, and $C_{2}(\cos \theta)$ are the three mBipoSH angular correlation functions that define the angular separation ($\theta$) space correlations corresponding to the NC beam nSI effect. Additionally, $f_{\lambda}(\hat n_1,\hat n_2)$ represents rank-2 tensor functions dependent on the directions of the measurement. The analytic form of these three angular correlation functions can be expressed using the expressions \eqref{eq8} and \eqref{qq} as
\begin{equation}
  C_\lambda(\cos \theta)=\sum_l \left(A_{l l}^{20} a_\lambda(l, l, 2, \cos \theta)+A_{l l+2}^{20} a_\lambda(l, l+2,2, \cos \theta) \right) \,\,\,\, \text{where } \lambda=0,1,2.
\end{equation}
The coefficients $a_\lambda$ can be easily calculated using equations \eqref{eq9} and \eqref{a_lambda coeffs}. The set of different   $a_\lambda$ coefficients can be expressed as

\begin{equation}
\begin{aligned}
    &a_1(l,l,2,\cos \theta)=(-1)^l\sqrt{\frac{10(2l+1)}{3l(2l+3)(2l-1)(l+1)}}\left((2l+3) P_l^1(\cos \theta)-2 P_{l+1}^2(\cos \theta)\right) \\
    &a_1(l,l+2,2,\cos\theta)=(-1)^l\sqrt{\frac{5}{9}\frac{1}{(2l+3)(l+2)(1+1)}}\times 2\left((2l+5) P_{l+2}^{1}(\cos \theta)- P_{l+3}^2(\cos \theta)\right)   \\ 
      &a_0(l,l,2,\cos \theta)=a_2(l,l,2,\cos \theta)=(-1)^l\sqrt{\frac{4(2l+1)}{(2l+3)(2l-1)(l+1)l}} P_l^2(\cos \theta)\\
        &a_2(l,l+2,2,\cos \theta)= (-1)^l\sqrt{\frac{2}{3}\frac{1}{(2l+3)(l+2)(l+1)}} P_{l+2}^2(\cos \theta) \\
        &a_0(l,l+2,2,\cos \theta)=a_2(l+2,l,2,\cos\theta)=(-1)^l\sqrt{\frac{2}{3}\frac{1}{(2.l+3)(l+2)(l+1)}} P_{l}^2(\cos \theta)
\end{aligned}
\end{equation}

The analytic forms for the three distinct mBipoSH  angular correlation functions are plotted for the $L=2$ non-circular (NC) elliptical beam case, as shown in the figure below. It is deduced from the plots that the nSI effect associated with the NC beam can be effectively captured by a set of three isotropic angular correlation functions in the angular separation ($\theta$) space. This results in noticeable signatures in the correlation functions, assuming a PT scan. We intend to utilize the concept of generalized scanning in future studies.

\begin{figure}
  \centering

\includegraphics[width=1\textwidth]{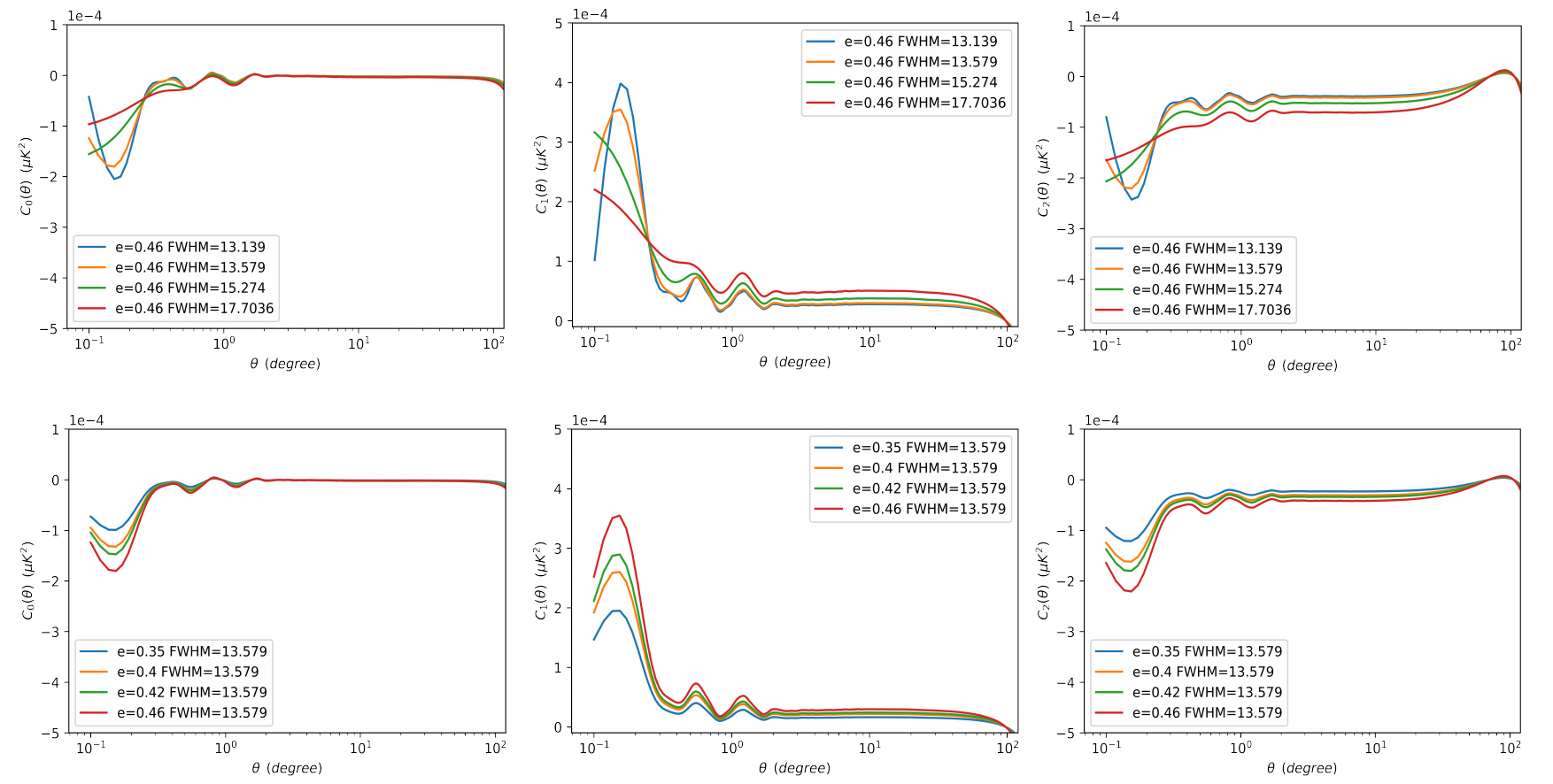}
\caption{Illustration of the mBipoSH isotropic correlation functions for the non-circular elliptical Gaussian beam. Top panel shows how the correlation functions vary with the change in $\theta_{\text{FWHM}}$. The bottom panel shows how these isotropic correlation functions change with respect to $e$.}
\end{figure}

\section{Discussions and Conclusions}\label{Conclusion}
In this paper, we extend the mBipoSH formalism to new Statistical Isotropy (SI) violation features in the CMB data, specifically the non-circular (NC) beam and cosmic hemispherical asymmetry (CHA). The mBipoSH functions represent the angular correlation function in the angular separation (real) space, providing a measure of correlations across points at various angular separations. We construct and plot the mBipoSH functions to analyze both CHA and NC beam effects. Our results show that CHA $(L = 1)$ can be explained by two mBipoSH correlation functions, while NC beam $(L = 2)$ is explained by a set of three angular correlation functions. This approach could be especially valuable for partial-sky observations, as it reveals SI violation effects in terms of real-space observables.

Our analysis of CHA indicates that distinguishing between CHA and random realizations of the CMB sky remains challenging. At this stage, the evidence does not decisively confirm or reject the presence of this asymmetry. In the case of the NC beam effect, we examined how angular correlations vary with respect to instrumental parameter settings, observing that changes in these parameters significantly impact the correlation structure. These findings suggest that the mBipoSH formalism holds strong potential for future missions equipped with high-resolution, partial-sky CMB maps. By applying this formalism, such missions could maximize insights into the statistical isotropy of large-scale structures and provide a clearer understanding of SI violation features in the CMB data.

\vskip20pt
\textbf{Acknowledgements:}
D.G. acknowledges CSIR and Raman Research Institute for providing the financial support  as 
Senior Research Fellow. DG would like to thank Rajorshi Chandra, Sarvesh Kumar Yadav and Shabbir Shaikh for helpful
discussions. The present work is carried out using the Computing facilities at RRI.

\nocite{*}
\appendix
\renewcommand{\theequation}{\Alph{section}.\arabic{equation}} 
\setcounter{equation}{0} 

\section*{Appendix} 
\label{appendix}

We review the derivation of the new mBipoSH angular correlation function in this appendix. We begin with the expansion of bipolar harmonics using arbitrary $l_1$ and $l_2$, creating a superposition of a small number of harmonics defined by the equation:
\begin{equation}\label{eq3}
\mathcal{Y}^{k}_{LM} (\hat n_1, \hat n_2) = Y^{L - k, k}_{LM} (\hat n_1, \hat n_2),\ \text{where}\ k = 0, 1, ..., L.\tag{A.1} 
\end{equation}
These new set of harmonics share the same rank $L$ but have the smallest possible ranks for their internal tensors, and their sum equals $L$. The basis composed of set of harmonics given by $\mathcal{Y}^{k}_{LM}$  represent the simplest bipolar harmonics of rank $L$ and can be readily analyzed.
From this, we can infer that, in general, the entire set of BipoSH can be reduced to these $L+1$ functions in this new basis such that internal ranks of this are limited to $L$. All other combinations of internal ranks beyond the value of $L$  can be formed using combining  these defined $L$  vectors, constructed in the appropriate manner.

In order to establish the relationship between any general bipolar harmonic basis element in terms of our reduced minimal harmonics basis, we use the concept of vector differentiation \cite{VMK}. Given that we are dealing with spherical harmonics we use the orbital angular momentum operators to do the mathematical analysis. 

For example, we can express the Legendre polynomial in the form of BipoSH as follows

\begin{equation}
P_l\left(\hat n_1 \cdot \hat n_2\right) = \frac{4 \pi}{2l+1} \sum_m Y_{lm}(\hat n_1) Y_{lm}^*(\hat n_2) = (-1)^l 4\pi Y^{ll}_{00}(\hat n_1, \hat n_2).\tag{A.2} 
\end{equation}
When we apply the orbital angular momentum operator $\hat{l}=-\mathrm{i}[r \times \nabla]$ to the Legendre polynomials, it gives the following relation
\begin{equation}
\hat{\boldsymbol{l}} P_l\left(\hat n_1 \cdot \hat n_2\right)=-\mathrm{i}\left[\hat n_1 \times \hat n_2\right] P_l^{(1)}\left(\hat n_1 \cdot \hat n_2\right)=4 \pi(-1)^l \sqrt{\frac{l(l+1)}{3(2 l+1)}} Y_{10}^{l l}\left(\hat n_1, \hat n_2\right).\tag{A.3} 
\end{equation}
From here, we can naively see that the nabla operator works as a raising and a lowering operator for bipolar harmonics. We can use this to create a complete basis for our new reduced minimal basis functions.

The tensor product of similar angular momentum operators can be used for building the bipolar harmonic basis. The angular momentum operator for rank L is denoted as $\{{\hat{\boldsymbol{l}}\}}_{LM}$. Using the notation of angular momentum operators and introducing the subsequent nabla operator from the angular momentum operator, the raising operator for spherical harmonics follows as

\begin{equation}
\left\{\{\nabla\}_k \otimes r^l Y_{lm}(\boldsymbol{r} / r)\right\}_L=\delta_{L, l-k}(-1)^k \frac{2 l+1}{2 l-2 k+1}\left[\frac{(2 l) !}{2^k(2 l-2 k) !}\right]^{\frac{1}{2}} r^{l-k} Y_{l-k, m}(\boldsymbol{r} / r). \tag{A.4} 
\end{equation}
The utilization of \(n = \frac{\boldsymbol{r}}{r}\) and \(n' = \frac{\boldsymbol{r'}}{r'}\) is to optimize the nabla operator, which operates on derivatives with respect to radial variables.
So, the corresponding rank decreasing operator  is
$$
\hat{\mathcal{O}}_{k q}^{l-}(r, \nabla)=(-1)^k \frac{2 l-2 k+1}{2 l+1} \sqrt{\frac{2^k(2 l-2 k) !}{(2 l) !}} \frac{1}{r^{l-k}}\{\nabla\}_{k q} r^l.
$$
Similarly, the rank increasing operator for the spherical harmonics is
$$
\hat{\mathcal{O}}_{k q}^{l+}(r, \nabla)=(-1)^k \sqrt{\frac{2^k(2 l+2 k+1)(2 l) !}{(2 l+1)(2 l+2 k) !}} r^{l+k+1}\{\nabla\}_{k q} \frac{1}{r^{l+1}} .
$$

Together, these operators can be simply written as
$$
\left\{\hat{\mathcal{O}}_k^{l \pm} \otimes Y_l(\boldsymbol{r} / r)\right\}_{j m}=\delta_{j, l \pm k} Y_{l \pm k, m}(\boldsymbol{r} / r)
$$
This expression helps to construct the relationship involving bipolar harmonics
\begin{equation}
    \begin{aligned}
\left\{\left\{\hat{\mathcal{O}}_{p}^{l-}(r_1, \nabla_1)\right.\right. & \left.\left.\otimes\hat{\mathcal{O}}_{q}^{l+}\left(r_2, \nabla_2\right)\right\} \otimes\left\{Y_l(\boldsymbol{r_1} / r_1) \otimes Y_l\left(\boldsymbol{r_2} / r_2 \right)\right\}\right\}_{L M} \\
= & \Pi_{L, l-p, l-q}\left\{\begin{array}{ccc}
q & p & L \\
l & l & 0 \\
l-q & l-p & L
\end{array}\right\} Y_{LM}^{l-k, l+q}\left(\boldsymbol{r_1} / r_1, \boldsymbol{r_2} / r_2\right) .
\end{aligned}\tag{A.5} 
\end{equation}
This expression serves as the rank increasing/decreasing operator for the BipoSH where  $\Pi_{a, b, \ldots}=\sqrt{(2 a+1)(2 b+1) \ldots}$  and the expression within the brackets represents a $9j$ symbol term.

Utilizing these raising and lowering operators defined for the framework of BipoSH basis, we can formally represent a general bipolar harmonic in terms of the action of the gradient operator tensor on the Legendre polynomials as 

$$
Y_{j m}^{l l^{\prime}}\left(\hat n_1, \hat n_2\right)=(-1)^{j+q} \frac{\sqrt{(2 l+1)\left(2 l^{\prime}+1\right)}}{4 \pi(2 n+1) !} c_j\left(l, l^{\prime}\right)\left\{\{\nabla\}_{k+\lambda_p} \otimes\left\{\nabla^{\prime}\right\}_{j-k}\right\}_{j m} r^q r^{\prime q} P_q\left(\boldsymbol{r} \cdot \boldsymbol{r}^{\prime} / r r^{\prime}\right) .
$$
Here, $c_j(l, l^{\prime})$ can be expressed as

\[
c_j(l, l^{\prime}) = \left[\frac{2^{j+\lambda_p}(l+l^{\prime}+j+1)!(2j+\lambda_p+1)!(l+l^{\prime}-j)!}{(j+\lambda_p+l-l^{\prime})!(j+\lambda_p+l^{\prime}-l)!}\right]^{1/2},
\]

 $q=\frac{1}{2}\left(l+l^{\prime}+j+\lambda_p\right), k=\frac{1}{2}\left(l-l^{\prime}+j-\lambda_p\right)$, and the parameter $\lambda_p$ characterizes the parity properties inherent in the spherical harmonics.


Consequently, utilizing the above defined relationships, we derive the subsequent reduction formulas for the expansions of any bipolar harmonic in the form of the minimal harmonic basis.

\begin{equation}\label{eq4}
      Y^{l_1,l_2}_{LM} (\hat n_1, \hat n_2) = \sum_{\lambda=\lambda_p}^{L} a^{\lambda_p}_{\lambda}(l_1,l_2,L,\cos\theta)  Y^{\lambda,L+\lambda_p-\lambda}_{LM} (\hat n_1, \hat n_2),\tag{A.6} 
  \end{equation}
  \noindent where the parameter $\lambda_p$ characterizes the space inversion property that conveys information about the parity of BipoSH functions expressed as
 \begin{equation}
        \lambda_p =\left\{
             \begin{array}{ll}
                0 &   \text{for even $(l_1 + l_2 - L)$ }\\
                1 &   \text{for odd  $ (l_1 + l_2 - L)$ },
             \end{array}
             \right.\tag{A.7} 
  \end{equation}
indicates that the BipoSH function $Y^{l_1l_2}_{LM}$ behaves as a tensor for even parity and a pseudo-tensor for odd parity and $\cos(\theta)$ is defined as dot product of the directions $ {n}_1 \cdot {n}_2$
.

In the equation \eqref{eq9}, the coefficients $a^{\lambda_p}_s$ can be explicitly expressed as
\begin{equation} \label{a_lambda coeffs}
\begin{aligned}
a_s^{\lambda_p}=\mathrm{i}^{\lambda_p}(-1)^{l^{\prime}} & {\left[\frac{2^j(2 l+1)\left(2 l^{\prime}+1\right)(2 j+1) !\left(l+l^{\prime}-j\right) ! s !\left(j-s-\lambda_p\right) !}{q!s\left(j-l+l^{\prime}\right) !\left(j-l^{\prime}+l\right) !(2 s+1) ! !\left(2 j-2 s-2 \lambda_p+1\right) ! !}\right]^{\frac{1}{2}} } \\
& \times \sum_{t=0}^{t_{\max }}(-1)^{s+t}\left(\begin{array}{c}
\frac{1}{2}\left(j^{\prime}-l\right) \\
t
\end{array}\right)\left(\begin{array}{c}
j-\lambda_p-t \\
s
\end{array}\right) \frac{\left(q-\lambda_p\right) ! !}{\left(q-\lambda_p-2 t\right) ! !} P_{j^{\prime}-t-s}^{(j-t)}(\cos \theta) 
\end{aligned}\tag{A.8} 
\end{equation}

with $q = l+l^{\prime}+j+1$, $j^{\prime} = j+l^{\prime}-\lambda_p$, $t_{\max }=\min \left\{j-\lambda_p-s ; \frac{1}{2}\left(j^{\prime}-l\right)\right\}$, and $\left(\begin{array}{l}m \\ n\end{array}\right)$ represents the binomial coefficient. The polynomials $P_n^{(m)}$ follow the symmetry relation given as $P_n^{(m)}(x)+(2 n-3) P_{n+1}^{(m-1)}(x)-P_{n+2}^{(m)}(x)=0$, Upon examining the equation \eqref{a_lambda coeffs}, it becomes apparent that the necessary symmetry for transposing \(l\) and \(l'\) is absent, thereby deviating from the definition of bipolar harmonics. This observation suggests that the equation does not represent the comprehensive form for \(a_\lambda\). Consequently, we apply these equations exclusively only when \(l\) is less than \(l'\), corresponding to coefficients \(s \geq \frac{L}{2}\). For the remaining coefficients, we employ a symmetry relation given as  \begin{equation}\label{eq6}
         a_{\lambda}(l_1,l_2,L,\cos\theta) =a_{L-\lambda_p-\lambda}(l_2,l_1,L,\cos\theta).\tag{A.9}  
     \end{equation} to determine the exhaustive set of minimal bipolar harmonics. This yields the complete set of minimal bipolar harmonics in an equivalently reduced form, satisfying the requirements of the completeness property of bipolar harmonics.


\label{Bibliography}

\bibliographystyle{JHEP}
\bibliography{ref}
\end{document}